\documentclass[aps,prl,showpacs,twocolumn,superscriptaddress,floatfix,fleqn]{revtex4}
\usepackage{graphicx}
\usepackage{bm}
\usepackage{amsmath,amssymb}

\renewcommand{\Im}{\mathop{\rm Im}}
\renewcommand{\Re}{\mathop{\rm Re}}

\begin{document}

\title{Mirrorless Negative-index Parametric Micro-oscillator }
\author{Alexander K. Popov}
\email{apopov@uwsp.edu}
\affiliation{Department of Physics \& Astronomy, University of Wisconsin-Stevens Point,
Stevens Point, WI 54481, USA.}
\author{Sergei A. Myslivets}
\affiliation{Institute of Physics of the Russian Academy of Sciences, 660036 Krasnoyarsk,
Russian Federation}
\author{Vladimir M. Shalaev}
\affiliation{Birck Nanotechnology Center and School of Electrical and Computer
Engineering, Purdue University, West Lafayette, IN 47907}
\date{\today}

\begin{abstract}
The feasibility and extraordinary properties of mirrorless parametric
oscillations in strongly absorbing negative-index metamaterials are shown.
They stem from the backwardness of electromagnetic waves inherent to this
type of metamaterials.
\end{abstract}
\maketitle

%\preprint{APS/123-QED}

% Force line breaks with \\

%\altaffiliation




%\date{February 12, 2003}


%Use showkeys class option if keyword
Optical negative-index (NI) metamaterials (NIMs) form a novel class of
artificial electromagnetic media that promises revolutionary breakthroughs
in photonics (see for example \cite{Sh}). Unlike ordinary positive-index
(PI) materials (PIMs), the energy flow and wave vector are counter-directed
in NIMs, which determines their extraordinary linear and nonlinear optical
(NLO) properties. Strong absorption is fundamentally inherent to NIMs, which
presents a severe detrimental factor toward their applications. Herein, we
show the feasibility and extraordinary properties of mirrorless generation
of entangled counter-propagating left- and right-handed photons in a
strongly absorbing micro sample of a NIM.

Frequency-degenerate multi-wave mixing and self-oscillations of
counter-propagating waves in ordinary materials have been extensively
studied because of their easily achievable phase matching. Phase matching
for three-wave mixing (TWM) and four-wave mixing (FWM) of contra-propagating
waves that are far from degeneracy  seem impossible in ordinary materials.
The possibility of TWM with mirrorless self-oscillations from
two co-propagating waves with nearly degenerate frequencies that
fall within an anomalous dispersion frequency domain and a far-infrared
difference-frequency counter-propagating wave in an anisotropic crystal was
proposed in \cite{Har} (and references therein) more than 40 years ago (see
also \cite{Yar}). However, far-infrared radiation is typically strongly
absorbed in crystals. For the first time, TWM backward-wave (BW) mirrorless
optical parametrical oscillator (BWMOPO) with all three different optical
wavelengths was recently realized \cite{Pas}. Phase-matching of
counter-propagating waves has been achieved in a submicrometer periodically
poled NLO crystal, which has become possible owing to recent advances in
nanotechnology. Both in the proposal \cite{Har} and in the experiment \cite%
{Pas}, \emph{the opposite orientation of wave vectors} was required for
mirrorless oscillations due to the fact that PI crystals were implemented.
As outlined, a major technical problem in creating BWMOPO stems from the
requirement of phase matching with the opposite orientation of wave vectors
in PIMs.
Herein, we show the feasibility of creating distributed feedback and BWMOPO
where \emph{an antiparallel orientation of wave vectors
of the coupled waves is not required anymore}. Thus BWMOPO at appreciably different
frequencies becomes possible while all the wave vectors of the coupled waves
remain co-directed. Such an opportunity makes phase matching much easier and
is offered by the backwardness of electromagnetic waves, which is natural to
NIMs. We note that NLO in NIMs still remains a less developed area of
electromagnetism. Two options are proposed in this work. One is TWM BWMOPO,
which implements intrinsic $\chi ^{(2)}$ nonlinearities associated, for
example, with the asymmetry of current-voltage characteristics from the
building blocks of NIMs \cite{Kiv}. A recent demonstration of the exciting
possibilities to craft NIMs with NLO responses that exceed those
from natural crystals is reported in \cite{Kl}. This option allows broad
frequency tunability; however, there might be technical problems to overlap
the NI, phase matching and strong NLO response frequency domains. The second
option, FWM BWMOPO, offers independent engineering of a strong nonlinearity $\chi ^{(3)}$ through embedded resonant NLO centers. In the vicinity of
resonances, it
%the nonlinearity
becomes exceptionally strong. In addition, the optical
properties of the composite become tailored by the means of quantum control.
\begin{figure}[!h]
\begin{center}
\includegraphics[width=.8\columnwidth]{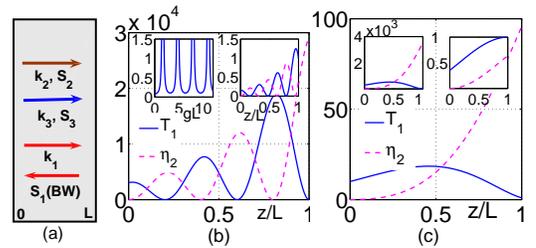}
\end{center}
\caption{Coupling geometry for three-wave BWMOPO, (a); the tailored
distribution of the signal, $T(z)=\left\vert {h_{1}(z)}/{h_{1L}}\right\vert
^{2}$, and of the idler, $\protect\eta_2=|E_2/E_1(L)|^2$, along the slab at $%
\Delta k=0$, (b) and (c). (b) and main plot (c): $\protect\alpha_{1}L=2.5$, $%
\protect\alpha_{2}L=-2$. (b): main plot: $gL=7.865$; upper right inset:
four-order decrease in $T_1(z)$ outside the geometrical resonance at $gL=10$%
; upper left inset: geometrical resonances in output signal $T_1(z=0)$. (c):
$gL=1.554$. Upper left insert: $\protect\alpha_2L=-\protect\alpha_1L=-2.5$.
Upper right insert: $\protect\alpha_2L=\protect\alpha_1L=2.5.$}
\label{f1}
\end{figure}

Figure \ref{f1}(a) depicts the coupling geometry for the proposed TWM
BWMOPO.
%%%%%%%%%%%%
We assume the wave at $\omega_{1}$ with the wavevector $\mathbf{k}_1$
directed along the $z$-axis to be a NI ($n_{1}<0$) signal. Therefore, it is
a backward wave  because its energy-flow $\mathbf{S}_{1} =(c/4\pi)[%
\mathbf{E_1}\times \mathbf{H_1}]$ appears directed against the $z$-axis. The
medium is illuminated by a higher-frequency PI wave at $\omega_{3}$
traveling along the $z$-axis ($n_{3}>0$). The two coupled waves with
co-directed wave vectors $\mathbf{k}_{3}$ and $\mathbf{k}_{1}$ generate a
difference-frequency idler at $\omega_{2}=\omega_{3}-\omega_{1}$, which is
also assumed to be a PI wave ($n_{2}>0$). The idler contributes back into
the wave at $\omega_{1}$ through TWM and thus enables optical parametric
amplification (OPA) at $\omega_{1}$ by converting the energy of the control
field at $\omega_{3}$ into the signal. Thus, all of the three coupled waves
have their wave vectors \emph{co-directed} along $z$, whereas the energy
flow of the signal wave, $\mathbf{S}_{1}$, is \emph{counter-directed} to the
energy flows of the two other waves, $\mathbf{S}_{2}$ and $\mathbf{S}_{3}$.
Such a coupling scheme is in strict contrast both with the conventional
phase-matching scheme for OPA in ordinary crystals, where all energy-flows
and phase velocities are co-directed, and with the TWM BWMOPO outlined
above, where the energy flow \emph{and wave vector} of one of the waves are
opposite to all others. The equations for slowly-varying amplitudes of the
signal and the idler take the form
\begin{eqnarray}
{dh_{1}}/{dz} &=& i\sigma_{1} h_{3}h_{2}^*\exp[i\Delta kz] +({\alpha_{1}}/{2}%
)h_{1},  \label{h1} \\
{dh_{2}}/{dz} &=& i\sigma_{2} h_{3}h_{1}^*\exp[i\Delta kz]-({\alpha_{2}}/{2}%
)h_{2},  \label{h2}
\end{eqnarray}
Here, $h_j$ are the magnetic components of the fields, $k_{j}=|n_j|%
\omega_j/c>0$; $\sigma_j=4\pi\chi^{(2)}{\epsilon_j\omega_j^2}/{k_{j}c^{2}}$;
$\Delta k=k_3-k_2-k_1$; $\chi^{(2)}$ is the magnetic nonlinear
susceptibility; $\alpha_{j}$ are absorption indices. The magnitude $h_{3}$
is assumed constant along the medium. Equation (\ref{h1}) exhibits \emph{%
three fundamental differences} as compared with TWM of co-propagating waves
in ordinary materials. First, there is an opposite sign of $\sigma_1$
because of $\epsilon_1<0$. Second, there is an opposite sign before $%
\alpha_{1}$ because the energy-flow $\mathbf{S_1}$ is directed against the
axis $z$. Third, the boundary conditions for $h_{1}$ are defined at the
opposite edge of the sample as compared to $h_{2}$ and $h_{3}$ because the
energy-flows $\mathbf{S_1}$ and $\mathbf{S_2}$ are counter-directed. The
transmission factor for the negative-index signal, $T(z)=\left\vert {h_{1}(z)%
}/{h_{1L}}\right\vert ^{2}$, is derived from the solution to the equations (%
\ref{h1}) and (\ref{h2}) as \cite{OL}
\begin{equation}
T_1(z=0)=T_{10}=\left|\dfrac{\exp \left\{-\left[ \left( \alpha_{1}/2\right)-s%
\right] L\right\}}{\cos RL+\left(s/R\right) \sin RL}\right|^2.  \label{T}
\end{equation}
Here, $s=({\alpha_{1}+\alpha_{2}})/({4})-i({\Delta k}/{2})$, $R=\sqrt{%
g^{2}-s^{2}}$ and $g=(\sqrt{\omega_1\omega_2}/\sqrt[4]{\epsilon_1\epsilon_2/%
\mu_1\mu_2}) ({8\pi}/{c}){\chi^{(2)}}h_{3}$. The fundamental difference
between the spatial distribution of the signal in ordinary and NI materials
is explicitly seen at $\alpha_j=\Delta k=0$. Then equation (\ref{T}) reduces
to
\begin{equation}
T_{10}=1/[\cos(gL)]^2,  \label{T1}
\end{equation}
whereas, in ordinary media, the signal would exponentially grow as $%
T_1\propto \exp(2gL)$. Equations (\ref{T}) and (\ref{T1}) present a sequence
of \emph{geometrical} resonances as functions of the slab thickness and of
the intensity of the control field $h_3$. Such extraordinary resonances
provide for the feasibility of attaining the \emph{oscillation threshold}
for the generation of \emph{the entangled counter-propagating left-handed, $%
\hbar\omega_1$, and right-handed, $\hbar\omega_2$, photons without a cavity}
at $g L \rightarrow (2j+1)\pi/2$.
A similar behavior is characteristic for
distributed-feedback lasers and is equivalent to a great extension of the NLO coupling length. It is known that even weak amplification per
unit length may lead to lasing provided that the corresponding frequency coincides with
a high-quality cavity or feedback resonances. The crucial role of the outlined geometrical
resonances, as well as the dramatic change caused by amplification of the
idler, is illustrated in Fig.~\ref{f1}(b),(c). The main plot in Fig.~\ref{f1}%
(b) shows the spatial behavior of the signal and the idler at a slightly
off-resonant value of $gL$, which results in the overall decrease of the
transmitted signal. Since the idler grows toward the back facet of the slab
at $z=L$ and the signal experiences absorption in the opposite direction,
the maximum of the signal for the given parameters appears closer to the
back facet of the slab. In general, the distribution of the signal along the
slab and, consequently, its output value depends strongly on the difference
of the absorption indices for the signal and the idler. As seen from the
comparison with the upper right inset, the signal maximum and the slab
transparency drop sharply with offset (here, an increase) in the magnitude
of $gL$ from its resonant value. The upper left inset in Fig.~\ref{f1}(b)
depicts the resonances in the output signal at $z=0$. A change in the slab
thickness or in the intensity of the control fields leads to significant
changes in the distributions of the signal and idler along the slab, which
is seen from a comparison of the main plots in Fig.~\ref{f1}(b) and (c).
Such a dependence is in stark contrast with its counterpart in
PI materials. The role of the additional amplification of the
idler provided by the control fields is readily seen from the comparison of
the upper left and right insets in Fig.~\ref{f1}(c).

\begin{figure}%[!h]
\begin{center}
\includegraphics[width=.8\columnwidth]{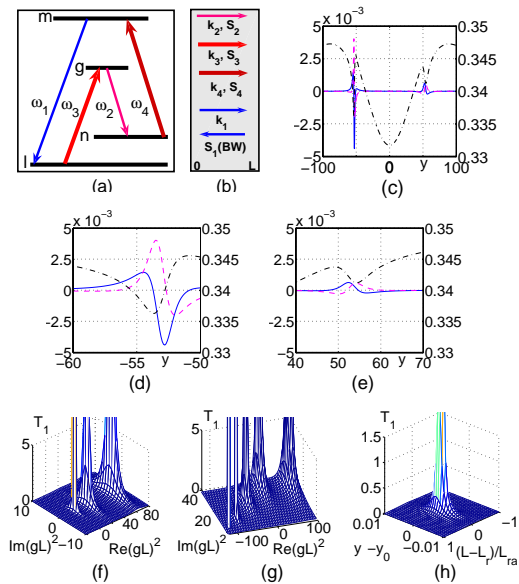}
\end{center}
\caption{Scheme of quantum-controlled resonant FWM BWMOPO, (a);
corresponding coupling geometry ($n_1<0$), (b); laser-induced nonlinear
interference structures and dispersion of local parameters, (c)-(e);
transmittance $T_{1}(z=0)$ of the NIM slab vs $\Re(gL)$ and $\Im(gL)$ at
different phase mismatch $\protect\delta kL$, (f) and (g); BWMOPO threshold
vs frequency resonance offset and the NIM slab thickness, (h). (c)-(e): $%
G_3=G_4=20$~GHz, $\Omega_3=-\Omega_4=\Gamma_{gl}$. $y=\Omega_1/\Gamma_{ln}$.
Solid: $\Im (g/\protect\alpha_{10})^2$; dash: $\Re(g/\protect\alpha_{10})^2$%
; dash-dot: $[(\protect\alpha_2+\protect\alpha_1+\protect\alpha_{NIM})/4%
\protect\alpha_{10}]^2$. (c): overview, (d): left resonance, (e): right
resonance. $\protect\alpha_2L=-2$, $\protect\alpha_1L=2.5$. (f): $\Delta k=0$. (g): $\Delta kL=-7\protect\pi$. (h): $\protect\delta k/\protect\alpha_{10}=0.248$, $L_r=32.6426 L_{ra}$, y$_0$=-53.13.}
\label{f2}
\end{figure}
The basic idea of resonant FWM BWMOPO is depicted in Fig. \ref{f2}(a),(b). A
slab of NIM is doped by four-level nonlinear centers [Fig. \ref{f2}(a)] so
that the signal frequency, $\omega_1$, falls in the NI domain ($n(\omega_1)<
0$), whereas all the other frequencies, $\omega_3$, $\omega_4$ and $\omega_2$, are in the PI domain. Here, we look for the possibility to achieve the
oscillation threshold for the BW at $\omega_1$ controlled by two lasers at $\omega_3$ and $\omega_4$. Due to FWM, these three fields generate an idler
at $\omega_2=\omega_3+\omega_4-\omega_1$, which may experience significant
amplification, either Raman or due to the population inversion at the
corresponding resonant transition, driven by the control fields. This opens
an \emph{additional channel} that may greatly enhance the coherent energy
transfer from the control fields to the signal. The amplified idler
contributes back to $\omega_1=\omega_3+\omega_4-\omega_2$ through FWM and
thus causes strongly enhanced parametric amplification of the signal. Unlike
ordinary off-resonant NLO, in this scenario many-orders of resonance
enhancement in the NLO coupling is accompanied by a \emph{strong change of
the local optical parameters} by the control fields. This is because the
control fields may produce a significant population transfer and even
inversion, as well as the modulation of the probability amplitudes.
Consequently, a split and other modifications of the resonance shapes occur
that stem from the quantum interference, which is shown in Fig. \ref{f2}%
(c)-(e). Switching between constructive and destructive interference may
serve as a tool for harnessing local optical coefficients. Alternatively,
such changes can be minimized, so that the major amplification would come
directly from the coherent energy transfer from the control fields to the
signal and the idler through the FWM coupling. The interference of quantum
pathways in the vicinity of the resonances may even lead to the fact that
the overall process ceases to be seen as a set of successive one- and
multi-photon elementary processes \cite{GPRA} (and references therein).
Similar to the case for TWM, the features of FWM in NIMs employed here
appear in stark contrast with those known for the ordinary PI materials due
to the fact that all wave vectors are co-directed in order to ensure maximum
phase matching but the energy flow $\mathbf{S_{1}}$ appears contra-directed
to $\mathbf{k}_4$ and, therefore to all other wave vectors [Fig. \ref{f2}(b)]. Like in the preceding case, this imposes extraordinary features on the
nonlinear propagation of the coupled waves. The slowly-varying electric amplitudes of
the coupled waves at $\omega_1$ and $\omega_2$ are given by the equations
\begin{eqnarray}
{dE_{1}}/{dz} &=&-i\gamma_1^{(3)} E_{2}^{\ast }\exp [i\Delta kz]+({\alpha_{1}%
}/{2} )E_{4},  \label{1} \\
{dE_{2}}/{dz} &=&i\gamma_2^{(3)} E_{1}^{\ast }\exp [i\Delta kz]-({\alpha_{2}}%
/{2 })E_{2}.  \label{2}
\end{eqnarray}
Here, $\gamma_{1,2}^{(3)}= ({4\pi}|\mu_{1,2}|\omega^2_{1,2}/k_{1,2}c^2)%
\chi_{1,2}^{(3)}E_{3}E_{4}$ are NLO coupling coefficients; the dielectric
permittivity and magnetic permeability, $\epsilon_j$ and $\mu_j$, are
negative at $\omega_1$; $\Delta k=k_{3}+k_{4}-k_{1}-k_{2}$; $\alpha_{j}$ are
the absorption or amplification coefficients; $\omega_{1}+\omega_{2}=%
\omega_{3}+\omega_{4}$; $k_{j}=|n_{j}|\omega_{j}/c>0$; and $\chi_{1,2}^{(3)}$
are effective electric nonlinear susceptibilities. The amplitudes of the
fundamental (control) waves $E_{3}$ and $E_{4}$ are assumed constant along
the slab. Equations (\ref{1}) and (\ref{2}) are similar to equations (\ref%
{h1}) and (\ref{h2}) and exhibit the same {fundamental differences} as
compared with their counterparts in ordinary PI materials: the sign of the
nonlinear polarization term $\gamma_{1}^{(3)}$ is opposite to that of $%
\gamma_{2}^{(3)}$, which occurs because $\mu_{1}<0$; the opposite sign
appears for $\alpha_{1}$; and the boundary conditions for $E_{1}$ must be
defined at the opposite side of the sample as compared to those for all
other waves. These differences lead to a counterintuitive evolution of the
signal and idler along the medium, similar to those discussed for the case
of TWM BWMOPO. With the aid of the solutions to equations (\ref{1}) and (\ref%
{2}) \cite{OLM}, the transmission (amplification) factor for the NI signal, $%
T_1(z)=\left\vert {E_{1}(z)}/{E_{1L}}\right\vert ^{2}$, is given by equation
(\ref{T}), where $g^2=\gamma_2^{\ast}\gamma_1$. A \emph{significant
difference} between the resonant and off-resonant NLO processes investigated
here is that the NLO susceptibilities and, therefore, the parameters $%
\gamma_1$ and $\gamma_2$ become complex and different from each other in the
vicinity of the resonances. Hence, the factor $g^2$ may become negative or
complex. This indicates an additional phase shift that causes \emph{further
radical changes in the nonlinear propagation features} which can be
tailored. Figures~\ref{f2}(c)--(e) show such laser-induced nonlinear
interference resonances in the coupling parameter $g^2$ with commensurate
real and imaginary parts. Figures~\ref{f2}(f),(g) prove that resonance
coupling and, hence, complex or negative magnitudes of $g^2$ may result in
the fact that \emph{phase matching $\Delta k=0$ ceases to be required}, and
the large phase mismatch $\delta k$ introduced by the host material can be
compensated through a frequency resonance offset for one of the fields. This
is in stark contrast with the optimization of OPA for a weak signal in the
off-resonant case for both ordinary materials and NIMs. Note that the
requirement of linear phase mismatch is known to optimize OPA for strong
fields due to the fact that the ratio of the intensities of the coupled
fields and, consequently, the coupling phase varies along the medium. A
strikingly similar behavior occurs for the weak signal and homogeneous
strong fields in resonant NIMs.
To demonstrate the outlined extraordinary NLO features and to prove the
proposed possibilities for producing the tailored transparency of NIMs and
FWM BWMOPO, we have adopted the following model for numerical simulations
presented in Fig.~\ref{f2}(c)--(e). The characteristic for molecules
embedded in a solid host were chosen as: energy level relaxation rates $%
\Gamma_n=20$, $\Gamma_g=\Gamma_m=120$; partial transition probabilities $%
\gamma_{gl}=7$, $\gamma_{gn}=4$, $\gamma_{mn}=5$, $\gamma_{ml}=10$ (all in $%
10^6$~s$^{-1}$); homogeneous transition half-widths $\Gamma_{lg}=1$, $%
\Gamma_{lm}=1.9$, $\Gamma_{ng}=1.5$, $\Gamma_{nm}=1.8$ (all in $10^{12}$~s$%
^{-1}$); $\Gamma_{gm}=5$, $\Gamma_{ln}=1$ (all in $10^{10}$~s$^{-1}$). We
assumed that $\lambda_2=756$ nm and $\lambda_4=480$ nm. The density-matrix
method described in \cite{GPRA} is used for calculating the
intensity-dependent local parameters while accounting for the quantum
nonlinear interference effects (NIE). This allows us to account for changes
in absorption, amplification, refractive indices and in magnitudes and signs
of NLO susceptibilities caused by the control fields. The changes depend on
the population redistribution over the coupled levels, which strongly
depends on the ratio of the partial transition probabilities. Figure~\ref{f2}%
(c)-(e) depicts the nonlinear interference structures in the local optical
parameters entering into Eq.~(\ref{T}) at the indicated control field
strengths and resonance detunings. Here, $\Omega_1=\omega_1-\omega_{ml}$;
other resonance detunings $\Omega_{j}$ are defined in a similar way.
Coupling Rabi frequencies are introduced as $G_{3}= E_3d_{lg}/2\hbar$, $%
G_{4} = E_4d_{nm}/2\hbar$. The value $\alpha_{NIM}$ denotes absorption
introduced by the host material, and $\alpha_{10}$ is the fully resonant
value of absorption introduced by the embedded centers at $\omega_1=\omega_{ml}$ with all driving fields turned off. With the given
parameters, the control fields cause essential changes in the level
populations, $r_l\approx0.487$; $r_g\approx0.484$; $r_n\approx0.014$; $%
r_m\approx0.014$ ($r_l+r_n+r_g+r_m=1$) which is followed by an appreciable
population inversion at the idle transition. Laser-induced NIE in the NLO
susceptibilities are followed by corresponding changes in absorption and
refractive indices. For the above indicated optical transitions, the
magnitude $G \sim 10^{12}$~s$^{-1}$ corresponds to control field intensities
of $I \sim$ 10 - 100 kW/(0.1mm)$^2$. As discussed above, the output signal
presents a set of distributed feedback-type resonances that provides for
BWMOPO. Optimization of the output signal at $z=0$ is determined by the
interplay of absorption, idler gain, FWM and by the wavevector mismatch.
This therefore appears to be a multi-parameter problem with sharp resonance
dependencies. The results of numerical analysis of Eq.(\ref{T}) for one such
transmission resonance based on the steady-state solutions to the density
matrix equations \cite{GPRA} is given in Fig.~\ref{f2}(h). The intrinsic
absorption of the host slab in the NI frequency domain has been assumed
equal to 90\%. In Fig.~\ref{f2}(h), we also introduce the slab length scaled
to the resonance absorption length, $L_{ra}=\alpha_{10}^{-1}$. The value of $%
L/L_{ra}$ is proportional to the product of the slab length and the number
density of the embedded centers. Figure~\ref{f2}(f) displays a narrow
geometrical resonance at $L=L_r=32.6427 L_{ra}$ for the optimum frequency
offset for the signal at $\Omega_1/\Gamma_{ln}=y_{0}$. This provides the
optimum compensation the given phase mismatch, $\delta k$, introduced by the
host material. Changes in $\Delta k$ introduced by molecules are accounted
for within the simulations. Fig.~\ref{f2}(h) shows that the width of the
oscillation resonance is on the scale of the narrowest (here Raman)
transition width and the resonant absorption length.
%The calculated optical parameters for
%$y=y_0$ are as follow:
%$\Im(\gamma_2/\alpha_{40})=-2.42\times10^{-2}$,
%$\Re(\gamma_2/\alpha_{40})=-5.61\times10^{-2}$,
%$\Im(\gamma_4/\alpha_{40})=2.94\times10^{-2}$,
%$\Re(\gamma_4/\alpha_{40})=-7.03\times10^{-2}$,
%$\Im(g^2/\alpha_{40}^2)=-3.35\times10^{-3}$,
%$\Re(g^2/\alpha_{40}^2)=3.22\times10^{-3} $, $\Delta
%k/\alpha_{40}=-0.165$, $\alpha_2/\alpha_{40}=-0.282$,
%$\alpha_4/\alpha_{40}=0.306$.
Amplification in the maxima in Fig.~\ref{f2}(h) reaches many orders of
magnitude, which indicates the feasibility of cavity-less oscillations.
Assuming a resonance
absorption cross-section $\sigma_{10}~\sim~10^{-16}$~cm$^2$, which is
typical for dye molecules, and a concentration of molecules $N~\sim~10^{19}$
cm$^{-3}$, we estimate $\alpha_{10}~\sim 10^3$~cm$^{-1}$ and the required
slab thickness to be in the microscopic scale $L~\sim (10 - 100) \mu$. The
contribution to the overall refraction index by the impurities is estimated
as $\Delta n < 0.5(\lambda/4\pi)\alpha_{10}\sim 10^{-3}$, which essentially
does not change the negative refractive index.

In conclusion, we propose mirrorless generation of a backward wave  and entangled counter-propagating left-handed signal and
right-handed idler photons in a
strongly absorbing microscopic samples of negative-index metamaterials. The
proposal implements the backwardness of electromagnetic waves inherent to
NIMs and coherent energy transfer from the control fields to the signal
through optical parametric coupling. One of the proposed schemes allows for
the independent engineering of a negative index and a strong
nonlinear-optical response in the metamaterial.

%\section*{Acknowledgments}
This work was supported by the U.~S. Army Research Laboratory and by the U. S. Army Research Office under grants number W911NF-0710261 and 50342-PH-MUR.

%\bibliography{PRL_NIM_MOL}% Produces the bibliography via BibTeX.

\end{document}